\documentclass[prd,twocolumn,showpacs,nofootinbib,preprintnumbers]{revtex4}

\usepackage{amsmath}
\usepackage{amsfonts}
\usepackage{graphicx}
\usepackage{dcolumn}

\def\be{\begin{equation}}
\def\ee{\end{equation}}
\def\ba{\begin{eqnarray}}
\def\ea{\end{eqnarray}}
\def\bs{\begin{subequations}}
\def\es{\end{subequations}}
\renewcommand{\S}{{\text{\tiny $\phi$}}}
\newcommand{\T}{{\text{\tiny $T$}}}
\newcommand{\eff}{{\text{\tiny eff}}}
\newcommand{\teta}{\tilde{\eta}}

\begin{document}

\title{Noncommutative models in patch cosmology}
\author{Gianluca Calcagni}
\email{calcagni@fis.unipr.it}
\affiliation{Dipartimento di Fisica, Universit\`{a} di Parma, Parco Area delle Scienze 7/A, I-43100 Parma, Italy\\}
\affiliation{INFN -- Gruppo collegato di Parma, Parco Area delle Scienze 7/A, I-43100 Parma, Italy}
\date{June 1, 2004}
\begin{abstract}
We consider several classes of noncommutative inflationary models within an extended version of patch cosmological braneworlds, starting from a maximally invariant generalization of the action for scalar and tensor perturbations to a noncommutative brane embedded in a commutative bulk. Slow-roll expressions and consistency relations for the cosmological observables are provided, both in the UV and IR region of the spectrum; the inflaton field is assumed to be either an ordinary scalar field or a Born-Infeld tachyon. The effects of noncommutativity are then analyzed in a number of ways and energy regimes.
\end{abstract}

\pacs{98.80.Cq, 04.50.+h, 11.10.Nx}
\preprint{\preprint{PHYSICAL REVIEW D \textbf{70}, 103525 (2004)  \hspace{9cm} hep-th/0406006}}

\maketitle


\section{Introduction}

The idea that the early Universe experienced a phase of accelerated expansion has come to a crucial point. Born as a panacea for some problems of the standard big bang scenario, the inflationary paradigm has been developed and refined during these years, always successfully explaining the available observational data. The upcoming generation of high-precision cosmological experiments such as WMAP (Wilkinson Microwave Anosotropy Probe) \cite{ben03,spe03,bri03} and Planck \cite{planc} might definitely operate a selection on the great amount of inflation-inspired models. On the other hand, new theoretical scenarios in which the high-energy physics grows more and more in importance have produced a set of interesting research fields implementing the traditional 4D cosmology: therefore we have string cosmology, braneworld cosmology, noncommutative cosmology, and so on.

In their seminal paper \cite{BH}, Brandenberger and Ho presented a model of large-scale perturbation spectra, in which a noncommutative geometrical structure is generated by the stringy spacetime uncertainty relation (SSUR)
\be \label{SSURph}
\Delta t \Delta x_p \geq l_s^2\,,
\ee
where $l_s$ is the string length scale and $x_p=a(t)x$ is a physical space coordinate. It has been argued that this is a universal property for string and brane theory \cite{yon87,LY,yon00}. This picture (henceforth ``BH") has then been further explored in \cite{HL1,TMB,HL2,HL3,KLM,KLLM,cai04} and presents many common features with trans-Planckian scenarios with a modified dispersion relation \cite{dan02,EGKS,MB,cre03,ACT,SS,KoKS}.

In this paper we construct another noncommutative model based on the same philosophy of \cite{BH} and confront it with  BH in its two versions, the first one with the Friedmann-Robertson-Walker (FRW) 2-sphere factored out in the action measure and the second one with a unique effective scale factor. Scalar and tensor amplitudes and indices as well as consistency equations are obtained through the slow-roll (SR) formalism, both for an ordinary scalar field and a Born-Infeld cosmological tachyon. All the observables turn out to be functions of a noncommutative parameter, called $\mu$, measuring the magnitude of the Hubble energy $H$ at the time of horizon crossing in comparison with the fundamental string mass $M_s=l_s^{-1}$.

Some works have studied the inflationary perturbations treating $\mu$ on either almost \cite{HL3,cai04} or exactly \cite{KLM,KLLM} the same ground of the SR parameters, computing UV amplitudes and indices via a double or SR expansion for small parameters, respectively. Here we will follow a different approach and consider $\mu$ as a distinct object with respect to the SR tower; we will keep only the lowest-SR-order part of the observables and regard any $\mu$-term as pertaining to these leading-SR-order quantities. We stress that, while the parameter $\mu$ accounts for nonlocal effects coming from the string scale $l_s$, the SR tower is determined by the dynamics of the cosmological inflationary expansion. Therefore, they describe quite distinct physical phenomena. In fact, there is no connection between $\mu$ and the recursively defined SR tower, although even the first SR parameter is introduced by hand; the elements of the tower are built up of time derivatives of $H$ and they all vanish in a de Sitter background, while $\mu$, which contains only the Hubble parameter and the string scale, does not. In particular, the lowest-SR-order spectral amplitudes, equivalent to those obtained in a quasi-de Sitter model, will depend on $\mu$.
Besides this motivation, such a procedure has additional advantages. For example, we can study regimes with not-so-small $\mu$ within the SR approximation; secondly, if one keeps the magnitude of $\mu$ unconstrained, one can also explore the IR region of the spectrum, $\mu \gg 1$, through appropriate techniques. 

These effective noncommutative models can be extended to braneworld scenarios in which the 3-brane experiences a cosmological expansion governed by an effective Friedmann equation. The precise theoretical setup is highly nontrivial even in the commutative case, because of the number of requirements to impose on the background forms and spacetime geometry in order to have a cosmological four-dimensional variety. We will phenomenologically assume to have a 3-brane in which the SSUR (\ref{SSURph}) holds for all the braneworld coordinates $\{x^\nu\}$, $\nu = 0,1,2,3$, while the extra dimension $y$ along the bulk remains decoupled from the *-algebra. 

A very qualitative way to see how such a noncommutative scenario might emerge is the following. One of the most promising braneworlds is the Randall-Sundrum (RS) model \cite{RSa,RSb} or its Gauss-Bonnet (GB) generalization (e.g., \cite{DLMS,cal3}, and references therein), motivated by M-theory as low-energy products of a dimensionally reduced 11D supergravity to a 10D string theory, down to a 5D effective gravity \cite{HW1,HW2,LOSW,LOW1,LOW2,KK,ADD} (see also \cite{MO}). The resulting 11D manifold is $\text{AdS}_5 \times X_{CY}$, where the brane is located at the fixed point $y=y_b$ of the $Z_2$ symmetry in the 5D anti-de Sitter bulk and the other six dimensions are compactified on a Calabi-Yau 3-fold $X_{CY}$. The 5D gravitational coupling is related to the 11D one by $\kappa_5^2=\kappa_{11}^2/V_{CY}$, where $V_{CY}$ is the internal volume of the Calabi-Yau space and $\kappa_{11}^2 = M_s^{-9}$; thus, we will identify the noncommutative string mass as the fundamental energy scale of the full theory. To diagonalize the noncommutative algebra and induce a pure 4D SSUR on the brane one might fix the expectation values of the 11D background fields such that the extra direction commutes, $[y,x^\nu]=0$. Some other subtleties to deal with are discussed in Sec. \ref{see}.

A useful approach to study perturbation spectra in braneworld scenarios is patch cosmology, which makes use of a nonstandard Friedmann equation, coupled to the slow-roll formalism; despite all the shortcomings of this approximated treatment of extra-dimensional physics, it gives several important first-impact informations. The four dimensional scenario is automatically included.

We will not provide a full derivation of standard results and leave the reader to consult the available literature.
For an introduction to inflation and perturbation theory, see \cite{lid97,MFB}. For a review on braneworld, see \cite{mar03}. For patch cosmology, scalar and tachyon inflation, and a more complete list of assumptions, technicalities, comments, and references, see \cite{cal3}.\footnote{To the reference list of \cite{cal3} we add \cite{NO,neu00,CNW,NOO1,LNO,NOO2,deh04,MS} for higher derivative and Gauss-Bonnet gravity models and \cite{PCZZ,PHZZ,GST,FK,GZ,KKKKL,DDK} for the cosmological tachyon.}

The plan of the paper is as follows. The general setup is established in Sec. \ref{setup}, further developing the formalism of \cite{cal3}; in Sec. \ref{models} we review the BH models and introduce a new prescription for the noncommutative action, describing then the UV and the IR spectral regions. Sec. \ref{zoo} fully develops these models; in Sec. \ref{disc} a detailed analysis summarizes the main results and Sec. \ref{rems} is devoted to concluding remarks.


\section{General setup} \label{setup}

We will keep the general framework of a noncommutative 3-brane in which, either in a limited time interval during its evolution or in a given energy patch, the cosmological expansion satisfies an effective Friedmann equation 
\be \label{FRW}
H^2=\beta_q^2 \rho^q\,,
\ee
where $q$ is constant and $\beta_q>0$ is some factor with energy dimension $[\beta_q]= E^{1-2q}$. Often we will use the more convenient parameter $\theta \equiv 2(1-q^{-1})$. Table \ref{table1} reports the characteristics of the 4D and braneworld cosmologies we will consider. There, $\kappa_5=8\pi m_5^{-3}=M_5^{-3}$ is the five-dimensional gravitational coupling, $\alpha$ is the Gauss-Bonnet coupling and $\lambda$ is the brane tension. Gravity experiments impose $M_5 \gtrsim 10^8$ GeV and $\lambda^{1/4} \gtrsim 10^3$ GeV; best-fit analyses of BH noncommutative models gives estimates for the string scale $M_s \sim 10^{11} - 10^{17}$ GeV \cite{HL1,TMB}. In typical Ho\v{r}ava-Witten scenarios, the fundamental scale is of order of the GUT scale, $M_s \sim 10^{16}$ GeV.
\begin{table}[ht]
\caption{\label{table1}The energy regimes described in the text.}
\begin{ruledtabular}
\begin{tabular}{ccdc}
Regime &   $q$   &   \theta  &          $\beta_q^2$          \\ \hline
GB     &  $2/3$  &      -1   & $(\kappa_5^2/16\alpha)^{2/3}$ \\
RS     &    2    &       1   &      $\kappa_4^2/6\lambda$    \\
4D     &    1    &       0   &         $\kappa_4^2/3$        \\
\end{tabular}\end{ruledtabular}
\end{table}

We neglect any contribution from both the Weyl tensor and the brane-bulk exchange; assuming there is some confinement mechanism for a perfect fluid with equation of state $p=w\rho$, the continuity equation on the brane reads $\dot{\rho}+3H (\rho+p) = 0$. Let us consider an inflationary four-dimensional flat universe filled with an homogeneous inflaton field $\psi$. If this is an ordinary scalar field $\phi(t)$ with potential $V$, then the energy density and pressure are
\be \label{rhop}
\rho = \frac{\dot{\phi}^2}{2} + V(\phi) = p+2V(\phi)\,,
\ee
and the effective equation of motion is
\be
\ddot{\phi}+3H \dot{\phi}+ V'=0\,.\label{eom0}
\ee
Another choice is to consider a homogeneous tachyon field $T(t)$, that is a scalar, causal field satisfying the Dirac-Born-Infeld action \cite{gar00,gib02}
\be \label{born}
S_\text{DBI}=\int d^4x\,(-a)^3V(T)\sqrt{1-\dot{T}^2}\,;
\ee
energy density and pressure read
\ba
\rho &=& \frac{V(T)}{c_S}\,,\label{Trho}\\
p &=& -V(T) c_S=-\frac{V(T)^2}{\rho}\,,
\ea
where $c_S=\sqrt{-w}=\sqrt{1-\dot{T}^2}$.


\subsection{Slow-roll parameters and commutative observables}

According to the inflationary paradigm, an early-Universe period of accelerated expansion is driven by a scalar field slowly ``rolling'' down its potential toward a local minimum. Thanks to this idea, one can construct a set of useful quantities (the SR parameters) which govern the dynamics of the system and make the computational task easier through suitable SR expansions. The first parameter is the time variation of the Hubble length $H^{-1}$,
\be \label{epsilon}
\epsilon \equiv -\frac{\dot{H}}{H^2}=\frac{3q(1+w)}{2}\,.
\ee
Full SR towers involving either the Hubble parameter or the inflaton potential can be defined from dynamical considerations within the Hamilton-Jacobi formulation of the problem. Here we will need only the first three bricks of the Hubble tower. For a scalar field, these are
\ba
\epsilon_\S &\equiv& \epsilon =\frac{3q\beta_q^{2-\theta}}{2} \frac{\dot{\phi}^2}{H^{2-\theta}}\,, \label{phepsilon}\\
\eta_\S     &\equiv&  -\frac{d \ln \dot{\phi}}{d \ln a}=-\frac{\ddot{\phi}}{H\dot{\phi}} \label{eta}\,,\\
\xi^2_\S    &\equiv&   \frac{1}{H^2} \left(\frac{\ddot{\phi}}{\dot{\phi}}\right)^. = \frac{\dddot{\phi}}{H^2\dot{\phi}}- \eta_\S^2\,.\label{xi}
\ea
The evolution equations of the parameters are second-SR-order expressions,
\bs \label{dotSR}
\ba
\dot{\epsilon}_\S &=& H\epsilon_\S \left[(2-\theta)\,\epsilon_\S-2\eta_\S\right]\,,\label{epsih'}\\
\dot{\eta}_\S     &=&  H\left(\epsilon_\S\eta_\S-\xi_\S^2\right)\,;\label{etah'}
\ea
\es
further time derivatives raise the SR order by one at each step. For the tachyon field, the SR parameters are
\ba
\epsilon_\T &\equiv& \epsilon = \frac{3q}{2} \dot{T}^2\,, \label{Tepsilon}\\
\eta_\T         &=&  -\frac{\ddot{T}}{H\dot{T}}\,, \\
\xi_\T^2  &=& \frac{1}{H^2} \left(\frac{\ddot{T}}{\dot{T}}\right)^. = \frac{\dddot{T}}{H^2\dot{T}}- \eta_\T^2\,,\label{Txi}
\ea
with time variation
\bs
\ba
\dot{\epsilon}_\T &=& -2 H \epsilon_\T \eta_\T\,,\label{Tdotepsi}\\
\dot{\eta}_\T &=& H\left(\epsilon_\T\eta_\T-\xi_\T^2\right)\,.\label{Tdoteta}
\ea
\es
Note that one can compute second-SR-order tachyon expressions by going to the formal limit $\theta \rightarrow 2$ in Eq. (\ref{epsih'}).

A derivation of the perturbation amplitudes will be seen in the noncommutative case. Here, we just quote the results for the commutative observables, denoted by a superscript $(c)$. The perturbation amplitudes can be written as
\be \label{ampls}
A^{(c)} \approx \frac{aH}{5\pi z}\,,
\ee
to lowest SR order (equivalently, in a quasi de Sitter 4D spacetime). In general, the squared function $z$ is
\be \label{zgen}
z^2 =\zeta_q \frac{a^2 (\rho+p)}{H^2}= \zeta_q\frac{a^2 (1+w)}{\beta_q^{2-\theta}H^\theta}\,,
\ee
where $\zeta_q$ is a proportionality coefficient; for an ordinary and tachyon scalar on the brane,
\ba
z(\phi) &=& \frac{a\dot{\phi}}{H}\,,\\
z(T)    &=& \frac{a\dot{T}}{c_S\beta_q^{1/q} H^{\theta/2}}\,,
\ea
with $\zeta_q(\phi)=1$ and $\zeta_q(T)=1/c_S$. The scalar amplitude is, to lowest SR order,
\be \label{Sdeg}
A_s^{(c)2} = \frac{3q\beta_q^{2-\theta}}{25\pi^2}\frac{H^{2+\theta}}{2\epsilon}\,,
\ee
where $\epsilon$ is given by either Eq. (\ref{phepsilon}) or (\ref{Tepsilon}), according to the model. The spectral index and its running are, to lowest SR order,
\ba
n_s^{(c)}-1     &\equiv& \frac{d \ln A_s^{(c)2}}{d \ln k}=2\eta_\S-4\epsilon_\S\,, \label{Sns}\\
\alpha_s^{(c)}  &\equiv& \frac{d n_s^{(c)}}{d \ln k}\approx \epsilon_\S\left[5(n_s^{(c)}-1)+4(3+\theta)\,\epsilon_\S\right],\nonumber\\ \label{Salp}
\ea
while for the tachyon
\ba
n_s^{(c)}-1          &=& 2\eta_\T-(2+\theta)\,\epsilon_\T\,, \label{Tns}\\
\alpha_s^{(c)}  &\approx& (3+\theta)\,\epsilon_\T\left[(n_s^{(c)}-1)+(2+\theta)\,\epsilon_\T\right].\label{Talp}
\ea
These equations can be obtained through Eq. (\ref{dotSR}) and the lowest-order relation
\be \label{dotk}
\frac{d}{d \ln k} \approx \frac{d}{Hd t}\,.
\ee
In Eqs. (\ref{Salp}) and (\ref{Talp}), the parameter $\xi^2$ has been dropped (see the discussion in \cite{cal3}).

Let $\chi$ be the inverse of the bulk curvature scale; the effective 4D Newton constant is $\kappa_4^2 = \kappa_5^2\chi/(1+4\alpha\chi^2)$, which in RS gives $\chi^2=\lambda\kappa_4^2/6$. The gravitational spectrum in RS and GB scenarios has been investigated in \cite{DLMS,LMW} for a de Sitter brane. It turns out that $A^{(c)2}_t=A_{t(\text{4D})}^{(c)2} F_\alpha^2 (H/\chi)$, where $A_{t(\text{4D})}^{(c)2}$ is the commutative 4D amplitude with $z_{\text{4D}}=\sqrt{2}a/\kappa_4$. $F_\alpha$ is a complicated function of $\alpha$ and $H/\chi$ determined by the normalization of the $y$-dependent part $\xi_0(y)$ of the graviton zero mode calculated on the brane position, $F^2_\alpha=\xi_0^2(y_b) \kappa_5^2/\kappa_4^2$. Writing $A^2_t=A_{t(\text{4D})}^2 z_{\text{4D}}^2/z^2$, one has
\be
z(h) = \frac{\sqrt{2}a}{\kappa_4 F_\alpha}\,.\label{zgrav}
\ee
This remarkably simple generalization of the 4D function $z$ is possible only because of the maximal symmetry of the de Sitter brane, which permits a variable separation of the wave equation for the Kaluza-Klein gravity modes, $h_{\mu\nu}(x,y) \rightarrow h_{\mu\nu}^{(m)}(x)\xi_m(y)$. To be consistent with the patch solution (\ref{FRW}),
we consider the approximated version $F_q^2$ of $F_\alpha^2$ in the proper energy limits: $F^2_1=1 \approx F^2_\alpha(H/\chi\!\!\ll\!\!1)$ in 4D, $F^2_2= 3H/(2\chi) \approx F^2_0(H/\chi\!\!\gg\!\!1)$ in RS, and $F^2_{2/3}= (1+4\alpha\chi^2)/(8\alpha\chi H)\approx F^2_\alpha(H/\chi\!\!\gg\!\!1)$ in GB \cite{DLMS}. We can write down the patch version of Eq. (\ref{zgrav}) by noting that in four dimensions the graviton background can be described by Eq. (\ref{zgen}) with $\zeta_1(h)=1$ and a perfect fluid $p_h= -\rho_h/3$ which does not contribute to the cosmic acceleration $\ddot{a}=aH^2(1-\epsilon)$, being $\epsilon=1$ in Eq. (\ref{epsilon}). Generalizing this stationary-solution trick one has $w(h)=2/(3q)-1$ and
\bs \label{zgrav2}
\ba 
z(h)  &=& \frac{\sqrt{2}a}{\kappa_4 F_q}\,,\\
F^2_q &\equiv& \frac{3q\beta_q^{2-\theta}H^\theta}{\zeta_q(h)\kappa_4^2}\,,
\ea
\es
where the coefficient $\zeta_q(h)$ is determined by the gravity model and is $\zeta_1(h)=1=\zeta_{2/3}(h)$ and $\zeta_2(h)=2/3$. The commutative tensor amplitude is then
\be \label{Tdeg}
A_t^{(c)2} = \frac{3q\beta_q^{2-\theta}}{25\pi^2}\frac{H^{2+\theta}}{2\zeta_q(h)}\,,
\ee
while the spectral index and its running are
\ba
n_t^{(c)} &\equiv& \frac{d \ln A_t^{(c)2}}{d \ln k}=-(2+\theta)\epsilon\,,\\
\alpha_t^{(c)}&=& (2+\theta)\,\epsilon\left[(n_s^{(c)}-1)+(2+\theta)\,\epsilon\right],
\ea
both for the ordinary scalar and the tachyon field. The tensor-to-scalar ratio is
\be
r^{(c)} \equiv \frac{A_t^{(c)2}}{A_s^{(c)2}} = \frac{\epsilon}{\zeta_q(h)}\,.
\ee


\subsection{Leading-order noncommutative observables}

Let $A$ denote a lowest-order perturbation amplitude, $A \in \{A_t,\,A_s\}$; in general, it can be written as
\be \label{Anoncom}
A(\mu_*,\,H,\,\psi) = A^{(c)} (H,\,\psi)\,\Sigma (\mu_*)\,,
\ee
where $\mu_*$ is a noncommutative parameter to be defined later, $A^{(c)}=A(\Sigma\!\!=\!\!1)$ is the amplitude in the commutative limit, and $\Sigma(\mu_*)$ is a function encoding leading-SR-order noncommutative effects. It will turn out that, up to $O(\epsilon^2)$ terms,
\be \label{dotsig}
\frac{d \ln \Sigma^2}{d \ln k} = \sigma \epsilon \,,
\ee
where $\sigma = \sigma(\mu_*)$ is a function of $\mu_*$ such that $\dot{\sigma}=O(\epsilon)$. The spectral index is
\be \label{spindex}
n \equiv \frac{d \ln A^2}{d \ln k} = n^{(c)}+\sigma\epsilon\,;
\ee
for the scalar spectrum, $n=n_s-1$. The index running is
\be
\alpha \equiv \frac{d n}{d \ln k} = \alpha^{(c)}+\frac{d^2 \ln \Sigma^2}{d \ln k^2}\,.
\ee
In the scalar field case, the last term can be written as
\be
\frac{d^2 \ln \Sigma^2}{d \ln k^2} = \sigma\epsilon_\S \left[\left(2-\theta-\bar{\sigma}\right)\epsilon_\S-2\eta_\S\right]\,,
\ee
[here $\bar{\sigma}= -\dot{\sigma}/(\sigma H\epsilon)$ to first SR order] with $\theta \rightarrow 2$ in the tachyon case. Because of Eq. (\ref{Anoncom}), the tensor-to-scalar ratio is $r=r^{(c)}$ and the consistency equations read
\ba 
\alpha_s(\phi) &=& r \zeta_q \left\{(5-\sigma)(n_s-1)\right.\nonumber\\
 &&+\left.\left[4(3+\theta)-\sigma (7+\theta+\bar{\sigma}-\sigma)\right]r \zeta_q\right\},\label{salph}\\
\alpha_s(T) &=& r \zeta_q \left\{(3+\theta-\sigma)(n_s-1)\right.\nonumber\\
&&+\left.\left[(2+\theta)(3+\theta)-\sigma (5+2\theta+\bar{\sigma}-\sigma)\right]r \zeta_q\right\}.\nonumber\\\label{talph}
\ea
The lowest-SR-order consistency equation for the tensor index is
\be \label{ntconeq}
n_t = [\sigma-(2+\theta)] \zeta_q r\,,
\ee
and its running is
\be
\alpha_t = r \zeta_q \left\{(2+\theta-\sigma)(n_s-1)+
\left[(2+\theta-\sigma)^2-\sigma \bar{\sigma} \right]r \zeta_q\right\}.\label{tenalph}
\ee
There is also a next-to-leading order version of Eq. (\ref{ntconeq}), which we will not consider here.


\section{Noncommutative models} \label{models}

Let us introduce the new time variable $\tau \in \mathbb{R}^+$, $\tau =\int a\,dt=\int da/H$. With a constant SR parameter $\epsilon$, an integration by parts with respect to $a$ gives
\be \label{tauusef}
\tau = \frac{a}{H} \frac{1}{1+\epsilon}\approx \frac{a}{H}\,.
\ee
Inequality (\ref{SSURph}) can be rewritten in terms of comoving coordinates as
\be \label{SSUR}
\Delta \tau \Delta x \geq l_s^2\,,
\ee
and the corresponding algebra of noncommutative spacetime is time independent, 
\be \label{alg}
[\tau,x]=il_s^2\,.
\ee
The *-product realizing Eq. (\ref{alg}) is defined as
\be \label{*}
(f*g)(x,\tau)=e^{-(il_s^2/2)(\partial_x\partial_{\tau'}-\partial_{\tau}\partial_{x'})}f(x,\tau)g(x',\tau')\big|_{\text{\tiny \begin{tabular}{l} $x'=x$ \\ $\tau'=\tau$ \end{tabular}}}\,.
\ee
This realization of noncommutativity is in contrast with
\be \label{alt*}
[x_\mu,x_\nu]=i\theta_{\mu\nu}\,,
\ee
where $\theta_{\mu\nu}$ is the noncommutative parameter. This type of noncommutative cosmology, which does not preserve the FRW symmetries, has been studied in \cite{CGS,LMMP,BG}. Other implementations can be found in \cite{LMMS,LMM,ABM,FKM,FKM2,GOR,BPN}.


\subsection{BH models}

In the following we will adopt the short notation $a=a(\tau)$ and $a_\pm\equiv a(\tau\pm kl_s^2)$. For the skipped details, see \cite{BH}. Consider now the action of a free scalar field $\Phi$ living in a (1+1)-dimensional FRW space. In the noncommutative models we will study, each conventional product is replaced by the *-product (\ref{*}); thus, the gravitational sector of the theory is not a completely passive spectator but is involved via the *-coupling of the metric with the matter content. The new 2D action reads, noting that $a^2=a*a$ \cite{BH},
\be \label{bhaction}
S_\text{BH}=\int d\tau dx\, \frac{1}{2}\left(\partial_\tau \Phi^\dagger*a^2*\partial_\tau \Phi-\partial_x\Phi^\dagger*a^{-2}*\partial_x\Phi\right).
\ee
In the comoving momentum space,
\be
\Phi(x,\tau)= {\cal V}^{1/2}\int_{k<k_0} \frac{dk}{2\pi}\,\Phi_k(\tau) e^{ikx}\,,
\ee
where ${\cal V}$ is the total spatial coordinate volume and 
\be \label{cutoff}
k_0 \equiv M_s a_\eff
\ee
is a cutoff realizing the stringy uncertainty relation. The most convenient way to recast the action is to write the scale factor as a Fourier integral, $a^2(\tau)=\int d\omega\, a_\tau^2(\omega) e^{i\omega\tau}$, and perform the *-products of the complex exponentials in the integrand, removing the cutoff in the limit $k_0 \rightarrow \infty$ when absorbing the $\delta$-integrals in momentum spaces. The result is
\be
S \approx {\cal V}\int_{k<k_0} d\tau dk\, \frac{1}{2}\left(\beta_k^+ \partial_\tau \Phi_{-k}\partial_\tau \Phi_k-\beta_k^-k^2\Phi_{-k}\Phi_k\right)\,,
\ee
where
\be \label{BHbeta}
\beta_k^\pm = \frac{1}{2} \left(a_+^{\pm 2}+a_-^{\pm 2}\right)\,.
\ee
Defining two new objects
\ba
a^2_\eff &\equiv& \sqrt{\frac{\beta^+_k}{\beta^-_k}}=a_+ a_-\,,\label{aeffbh}\\
y^2 &\equiv& \sqrt{\beta^+_k \beta^-_k}=\frac{a^2_+ + a^2_-}{2a_+a_-}\,,
\ea
and the effective conformal time coordinate
\be \label{teta}
\frac{d \teta}{d\tau}=a^{-2}_\eff\,,
\ee
the scalar action becomes
\be \label{1+1}
S \approx {\cal V}\int_{k<k_0} d\teta dk\, \frac{1}{2}y^2\left(\Phi'_{-k}\Phi'_k-k^2\Phi_{-k}\Phi_k\right)\,,
\ee
where the primes are derivatives with respect to $\teta$. 


\subsection{A new prescription for noncommutativity}

yclic permutations of the *-product inside the integral (\ref{bhaction}) leave the action invariant. Therefore, it is natural to see whether a different noncyclic ordering of the factors gives a theory with interestingly new predictions. The other nontrivial noncommutative action one can obtain is
\ba
S_\text{new} &=& {\cal V}\int d\tau dx\, \frac{1}{2}\left(\partial_\tau \Phi^\dagger*a*\partial_\tau \Phi*a\right.\nonumber\\
&&\left.\qquad\qquad\quad-\partial_x\Phi^\dagger*a^{-1}*\partial_x\Phi*a^{-1}\right).
\ea
The same computational pattern of the previous section leads to Eq. (\ref{1+1}) with $\beta_k^\pm$ given by
\be
\beta_k^\pm = \frac{a^{\pm 1}}{2} \left(a_+^{\pm 1}+a_-^{\pm 1}\right)\,,
\ee
and
\ba
a_\eff^2 &=& a \sqrt{a_+ a_-}\,, \label{aeffnew}\\
y^2 &=& \frac{a_+ + a_-}{2\sqrt{a_+a_-}}\,.
\ea
In this case there is only a partial smearing of the product of scale factors and one might guess that the resulting noncommutative phenomenology would be less pronounced than that of BH model. In the UV limit it will turn out that, within a given variation of the noncommutative parameter and in \emph{some} region in the space of parameters, the range of the quantity $|\alpha_s(\phi)-\alpha_s(T)|$ is slightly smaller than in the BH model but always of the same order of magnitude. In the infrared region, however, the two models are almost undistinguishable; see Sec. \ref{disc}.


\subsection{Four-dimensional effective actions and amplitudes} \label{see}

When going to 3+1 dimensions, the measure $z_k^2$ of the integral will contain the nonlocal effect coming from the SSUR:
\be \label{action}
S \approx {\cal V}\int_{k<k_0} d\teta d^3k\, \frac{1}{2}z_k^2\left(\Phi'_{-k}\Phi'_k-k^2\Phi_{-k}\Phi_k\right)\,.
\ee
Here we will consider two classes of models. In the first one, we suppose the total measure to be given by the product of the noncommutative (1+1)-measure and the commutative one:
\be \label{mod1}
z_k = z y\,;
\ee
then, as we are going to show in a moment,
\be \label{SigmaI}
\Sigma = \frac{a^2_\eff}{a^2 y} \qquad \mbox{(class 1)}\,.
\ee
These models, in which the FRW 2-sphere is factored out, will be dubbed as ``1.'' Another interesting prescription consists in replacing the commutative scale factor in $z_k$ with the effective one; then, $ay \rightarrow a_\eff$, 
\be \label{mod2}
z_k=z \frac{a_\eff}{a}\,,
\ee
and
\be\label{SigmaII}
\Sigma = \frac{a_\eff}{a}  \qquad \mbox{(class 2)}\,;
\ee
models with this $\Sigma$ will be named ``2.''

Let us now look at cosmological perturbations coming from an inflationary era and assume, as it is the case, that $\Phi$ is a generic perturbation satisfying the action (\ref{action}). The spectral amplitude coming from the $k$-th mode of the perturbation is
\be \label{ampli}
A^2 \equiv \frac{2k^3}{25\pi^2} \left\langle |\Phi_k|^2\right\rangle\!\Big|_*\,,
\ee
where angle brackets denote the vacuum expectation value and the expression is evaluated at the reference time $\teta_*$ to be discussed in a while. Via a change of variable, 
\be
u_k=-z_k\Phi_k\,,
\ee
the action (\ref{action}) gives the Mukhanov equation
\be \label{muk}
u_k''+\left(k^2-\frac{z_k''}{z_k}\right)u_k=0\,.
\ee
Noting that $d \teta /d\eta =(a/a_\eff)^2$, we get the useful relation
\be \label{useful}
\teta \approx \frac{-1}{aH}\left(\frac{a}{a_\eff}\right)^2\,,
\ee
in the lowest SR approximation. If the SR parameters are small, then they are constant to leading order because their derivatives are higher order. It is then possible to solve the Mukhanov equation with exactly constant SR parameters and perturb the obtained solution. Such cosmological solutions do exist and can be constructed in a variety of situations; among them, a particularly important one is power-law inflation, which we will use when considering the infrared region of the spectrum. Therefore,
\be
\frac{1}{z_k}\frac{d^2z_k}{d\teta^2} \approx \left(\frac{a_\eff}{a}\right)^4 \frac{1}{z}\frac{d^2z}{d\eta^2}=\left(\frac{a_\eff}{a}\right)^4\frac{\nu^2-1/4}{\eta^2}\approx\frac{\nu^2-1/4}{\teta^2}\,,
\ee
where $\nu = 3/2+O(\epsilon)$. With constant $\nu$, the solution of this equation is the same as that of the commutative case, namely $|u_k| \propto (-\teta)^{1/2} H_\nu^{(1)}(-k\teta)$, where $H_\nu^{(1)}$ is the Hankel function of the first kind of order $\nu$. In the long wavelength limit $k/(aH)\rightarrow 0$, when the mode with comoving wave number $k$ is well outside the horizon, the appropriately normalized solution becomes, from Eq. (\ref{useful}),
\be
|u_k|^2=\frac{1}{2k}\left(\frac{-1}{k\teta}\right)^2=\frac{1}{2k}\left(\frac{aH}{k}\right)^2\left(\frac{a_\eff}{a}\right)^4\,;
\ee
finally, one gets Eq. (\ref{Anoncom}) by inserting either definition (\ref{mod1}) or (\ref{mod2}) in Eq. (\ref{ampli}).

Given a noncommutative brane in a commutative bulk, the nonlocal smearing will only affect the pure four-dimensional part of the graviton-zero-mode action, while leaving the pure transversal normalization unchanged; from the discussion in Sec. \ref{setup}, it is then clear that the noncommutative tensor spectral amplitude will be $A_t^2 =A_t^{(c)2}\Sigma^2\propto \xi_0^2(y_b) A_{t(\text{4D})}^2$. Therefore,
for the gravitational spectrum, $\Phi$ denotes the coefficient functions of the noncommutative 4D polarization tensor $h_{\mu\nu}^{(0)}(*x)$ and $z$ is given by Eq. (\ref{zgrav2}).

In the case of the inflaton field, $\Phi={\mathcal R}$ is the curvature perturbation on comoving hypersurfaces, generated by quantum fluctuations of the field filling the early Universe.

The action and Mukhanov equation for a perturbation generated by a tachyon field has an additional factor in front of $k^2$ in Eqs. (\ref{action}) and (\ref{muk}), namely the speed of sound for the perturbation: $k^2 \rightarrow c_S^2k^2$ \cite{GM,FKS}. Since the SSUR does not affect products of homogeneous quantities, the noncommutative generalization of the tachyonic scalar amplitude is straightforward \cite{LL}. Now, one may ask how the inhomogeneous version of the original Born-Infeld action (\ref{born}) is modified when inserting the *-products. Let us recall that noncommutativity naturally arises in string theory when a Neveu-Schwarz--Neveu-Schwarz (NS-NS) $B$-field is switched on in the low-energy tree-level action. However, this results in a linearization of the tachyonic action and, on the other hand, a large noncommutative parameter may trigger brane decay processes \cite{DRM}; therefore, the simple noncommutative version of the cosmological tachyon might seem too na\"{i}ve.

Anyhow, tachyon scenarios are not new to counterintuitive behaviors. In the slow-roll approximation, $\epsilon \propto \dot{T}^2 \ll 1$, the action (\ref{born}) can be linearized and the rescaled field $\phi=\sqrt{V}T$ behaves like an ordinary scalar; nevertheless, the theoretical prediction encoded in the consistency relations is different with respect to that of the genuine scalar scenario [see Eqs. (\ref{salph}) and (\ref{talph})]. Here, something similar happens, imagining to turn on and increase the $B$-field smoothly, and the final result differs from the scalar case indeed. 

Moreover, the stringy linearization is a feature of realization (\ref{alt*}) rather than (\ref{alg}) and the former may give rise to a different cosmological model in which FRW isotropy is not preserved \cite{LMMP}; also, \textit{a priori} it would be highly nontrivial to construct a Lorentz-violating cosmological brane model (in fact, in the case of a de Sitter brane, maximal symmetry is crucial for coordinate-separating the graviton wave equation \cite{LMW,DLMS}).

To further understand the difficulties lying in a full implementation of noncommutative string theory in cosmology, it is important to stress that all that has been said about the algebra (\ref{alt*}) (i.e. instability and cosmological scenarios) is true only in a purely spatial *-product, $\theta_{0i}=0$. When trying to introduce noncommutativity in both space and time, as is the case of realization (\ref{alg}), it may be difficult to achieve a coherent, well-defined theory. In fact, in the Seiberg-Witten limit reproducing the noncommutative geometry, $\theta_{\mu\nu}$ and $\alpha'B_{\mu\nu}$ are kept fixed while $B_{\mu\nu}\rightarrow \infty$ and the Regge slope $\alpha' \rightarrow 0$. Let $E_i=B_{0i}$ be the electric part of the NS 2-form and assume $E=|E_i|\neq 0$. Then, while the $B$-field goes to infinity and approaches the critical value $E_{cr}=(2\alpha')^{-1}$, a classical instability develops and the rate of open string pair production diverges \cite{BP}; heuristically, the string is tore apart by the increasing electric field strength. 
For these reasons we regard algebra (\ref{alg}) as the starting point of the cosmological setup rather then the ultimate product of some high-energy theory, for the moment leaving the details of the latter aside. 


\subsection{The UV region}

In order to correctly evaluate the perturbation spectra, one must determine the time $\teta_0$ when the $k$-th mode is generated and, later, when it crosses the Hubble horizon. Because of the momentum cutoff (\ref{cutoff}), the analysis for the noncommutative case must be conducted separately in the mildly and strongly noncommutative regions. 

From the very beginning, one can define the time $\teta_*$ when a perturbation with wave number k crosses the horizon by the formula $k_* \equiv k(\teta_*) = a(\teta_*) H(\teta_*)$. This relation provides an operative definition of the number of e-foldings ($k \propto H \exp N$) and the time variation of $k$, Eq. (\ref{dotk}). Of course, this is valid for any cosmology in which time definitions have zero uncertainty, that is, for commutative cosmologies and noncommutative cosmologies in the range far from the upper bound (\ref{cutoff}), in the so-called ultraviolet region, where $k_* \ll k_0$. In fact, the time of horizon crossing is different from its commutative counterpart $\teta_c$, since $\teta_c < \teta_*$ and the crossing mode is delayed \cite{BH}. In \cite{KLLM} this effect is quantified as $k_c/k_* = \exp [-(\teta_*-\teta_c)]$.

On the contrary, one might define the horizon crossing through the $z$ function as
\be \label{optpivot}
k^2_* = \frac{z_k''}{z_k} \approx 2(aH)^2\,,
\ee
and get an extra factor of 2; due to the structure of the Mukhanov equation, this approach would be valid in any case, let it be the commutative or the noncommutative one. 

In the UV region, the cosmological energy scale when the perturbation is generated is much smaller than the stringy scale, $H(\teta\! >\!\!\teta_0) \leq H(\teta_0) \ll M_s$, and noncommutative effects are soft; thus, the smeared versions $a_\pm$ of $a$ can be approximated by $a$ since
\be \label{UV}
kl_s^2 \ll \tau_* \qquad \mbox{(UV region)}\,,
\ee
from Eq. (\ref{tauusef}). It is convenient to define the noncommutative parameter
\be \label{mu}
\mu \equiv \left(\frac{kH}{a M_s^2}\right)^2,
\ee
whose time derivative is
\be \label{dotmu}
\dot{\mu} = -4H\mu\epsilon\,.
\ee
Note that this relation states that $\mu$ is almost constant in a rapidly accelerating background, regardless of its magnitude. The analogy with the evolution equations of the SR tower, e.g., Eq. (\ref{dotSR}), suggested the authors of \cite{KLM,KLLM} treat $\mu$ as a sort of SR parameter, keeping all the parameters at the same truncation level in the expressions of the UV observables.

At horizon crossing,
\be \label{mu*}
\mu_*=\mu |_{k=\sqrt{2}aH}=2\left(\frac{H}{M_s}\right)^4\,,
\ee
and Eq. (\ref{dotmu}) is valid for $\mu_*$, too. The ultraviolet region is by definition the region in which $H/M_s \ll 1$;\footnote{Without risk of confusion, we will continue to use the symbol $\mu$ to indicate the ratio $H/M_s$ when discussing the UV limit ($\mu \ll 1$) of spectral quantities.} it is characterized by long wavelength perturbations generated inside the Hubble radius and, in a cosmic microwave background (CMB) spectrum, this would correspond to the portion of the Sachs-Wolfe (inflationary) plateau with not-too-small spherical modes, $10 \lesssim l \lesssim 100$.

In the commutative case, to use one pivot scale instead of the other amounts to different next-to-lowest-order expansions in the SR parameters; the 4D consistency equations are thus unaffected, since the introduction of the optimized pivot scale (\ref{optpivot}) results in a rescaled coefficient $C \rightarrow C+\ln \sqrt{2}$ and this one is not present in them (see, e.g., \cite{KLLM} and references therein for details). This is also true in the RS scenario \cite{cal2} as well as in general patch cosmology \cite{cal5}.

In the noncommutative case, the change of the pivot scale doubles the magnitude of the parameter (\ref{mu*}). The resulting models will display the same theoretical features of the $k=aH$ models, but shifted backward along the energy scale determined by the ratio $H/M_s$. Observational constraints should take the rescaling of the string mass into account, when changing the pivot scale.

In the limit (\ref{UV}), we can Taylor expand the scale factors $a_\pm$ around $\tau$ for small $k$. To first order in the SR parameters and to all orders in $\mu$, the nonlocal dependence of the scale factor is
\ba
a(\tau \pm kl_s^2) &=& a(\tau)\,\left\{1 \pm \sqrt{\mu}+\left[\pm\sqrt{\mu}\right.\right.\nonumber\\
&&- \left.\left.(1 \pm\sqrt{\mu}) \ln (1\pm\sqrt{\mu})\right]\epsilon\right\}+O(\epsilon^2)\,,\nonumber\\
\ea
where the factor in front of $\epsilon$ comes from a series whose radius of convergence is $\mu \leq 1$. More precisely, when $\mu_* \leq 1$ then $H/M_s \lesssim 0.8\,.$ Since we are interested in lowest-SR-order amplitudes, we can neglect the SR tower and find
\be \label{aUV}
a_\pm \approx \left(1 \pm \sqrt{\mu}\right)a\,.
\ee
The concrete procedure to compute the spectral amplitudes will be to use the horizon crossing formula (\ref{optpivot}) at $\teta_*$ in the UV region, and the saturation time $\teta_0$ in the IR region. In \cite{BH} and other papers these instants are dubbed $\teta_k$ and $\teta_k^0$, respectively, to highlight the dependence on the wave number.


\subsection{BH model IR region}

In the IR region things are quite different: the wave modes are generated outside the horizon and, since they are frozen until they cross the horizon, their magnitude depends on the time when they were generated. This corresponds to the ($k$-dependent) time $\teta_0$ when the SSUR is saturated, $k(\teta_0)=k_0(\teta_0)$, and quantum fluctuations start out with their vacuum amplitude. The effective and smeared scale factors must be evaluated at this instant; the expansion (\ref{aUV}) is no longer valid since $H \gg M_s$ in the infrared. To proceed one can explicitly use the exact solution around which the equation of motion for the perturbation has been expanded. The power-law solution corresponds to a constant index $w$, when the scale factor is $a(\tau)=\alpha_0 \tau^{n/(n+1)}$, and $H= n\alpha_0\tau^{-1/(n+1)}/(n+1)$. For an exponential scale factor (de Sitter expansion, $n \rightarrow \infty$), $a(\tau)= H\tau$, in accordance with Eq. (\ref{tauusef}). From equations (\ref{cutoff}) and (\ref{aeffbh}),
\be
\tau_0 = kl_s^2 \sqrt{1+\delta}\,,
\ee
where $\tau_0=\tau(\teta_0)$ and
\be
\delta \equiv \left(\frac{2}{\mu_*}\right)^{1/2} = \left(\frac{M_s}{H}\right)^2\,.
\ee
In the infrared region, 
\be \label{IR}
kl_s^2 \approx \tau_0 \qquad \mbox{(IR region)}\,,
\ee
and
\ba
a     &=& Hkl_s^2 \sqrt{1+\delta}\,,\label{adelta}\\
a_\pm &=& Hkl_s^2 \left(\sqrt{1+\delta} \pm 1\right)\,,\label{apmdelta}
\ea
where evaluation at $\tau_0$ is understood. When $\delta \gg 1$, we recover the UV or quasicommutative region since $kl_s^2 \ll \tau_0 \leq \tau_*$. Actually, the UV and IR spectra may be joined together in an intermediate region, as it was shown in \cite{TMB}; in particular, see their Eq. (12), corresponding in the de Sitter limit to $\Sigma^2 \sim \delta (1-3\sqrt{\mu_*/2})$. We will not be able to recover this spectrum within our formalism; however, we will describe other hybrid regimes by using the methods adopted in the IR region ($2 \leq l \lesssim 10$) for $\delta \gg 1$. For future reference, note that
\be \label{dotdelta}
\dot{\delta} =2\delta H\epsilon\,.
\ee


\subsection{New model IR region}

In the ``New'' model, the effective scale factor is given by Eq. (\ref{aeffnew}). From Eq. (\ref{cutoff}),
\be
\tau_0 = kl_s^2 \sqrt{1+\gamma}\,,
\ee
where
\be
\gamma \equiv \frac{1}{2} \left(\sqrt{1+4\delta^2}-1\right)\,.
\ee
With this definition, the new expressions for $a$ and $a_\pm$ are identical to Eqs. (\ref{adelta}) and (\ref{apmdelta}), with $\delta$ replaced by $\gamma$. Equation (\ref{dotdelta}) is replaced by
\be
\dot{\gamma} = \frac{4\gamma(\gamma+1)}{1+2\gamma}H\epsilon\,.
\ee
In the far IR region, $\gamma \approx \delta^2 \ll 1$, while in the UV limit $\gamma \approx \delta \gg 1$.

Without further justifications, the IR region of the spectrum, $H \gg M_s$, may be not very satisfactory from a string-theoretical point of view, both because we are above the fundamental energy scale\footnote{However, the space-momentum stringy uncertainty relation, implying $\Delta x_p \geq l_s$, is not a universal property of the theory.} and due to the above-mentioned classical instabilities. As it is done in many other occasions in early-Universe cosmology, we will turn a blind eye to this point and seek what are the observational consequences of the extreme regime of the present noncommutative models.


\section{Noncommutative zoology} \label{zoo}

We are ready to collect all the machineries developed so far and inspect the noncommutative models at hand. 


\subsection{BH1}

In the BH1 case, 
\be
\Sigma^2 = \frac{2(a_+a_-)^3}{a^4 (a^2_+ + a^2_-)}\,.
\ee
In the UV region,
\bs \ba
\Sigma^2     &=& \frac{(1-\mu_*)^3}{1+\mu_*}\,,\\
\sigma       &=& \frac{8\mu_*(2+\mu_*)}{1-\mu_*^2}\,,\\
\bar{\sigma} &=& \frac{8(\mu_*^2+\mu_*+1)}{(2+\mu_*)(1-\mu_*^2)}\,.
\ea\es 
For $\mu \ll 1$ one recovers the nearly commutative, $\mu$-expanded behavior\footnote{Throughout the paper we will keep only the leading-order term in the approximated $\bar{\sigma}$ since there is a $\sigma$ factor in front of it in Eqs. (\ref{salph}) and (\ref{talph}).} 
\bs \ba
\Sigma^2     &\approx& 1-4\mu_*\,,\\
\sigma       &\approx& 16\mu_*\,,\\
\bar{\sigma} &\approx&  4\,.
\ea\es 
In the IR region,
\bs 
\ba
\Sigma^2 		 &=& \frac{\delta^3}{(2+\delta)(1+\delta)^2}\,,\\
\sigma 			 &=& \frac{4(2\delta+3)}{(2+\delta)(1+\delta)}\,,\\
\bar{\sigma} &=& \frac{2\delta(2\delta^2+6\delta+5)}{(3+2\delta)(2+\delta)(1+\delta)}\,.
\ea
\es 
In the commutative limit ($\delta \gg 1$), $\Sigma^2 \approx 1$, while in the strongly noncommutative regime ($\delta \ll 1$), $\Sigma^2 \approx \delta^3/2$ and $\sigma \approx 6-5\delta$, in agreement with \cite{TMB}.\footnote{Eqs. (44)--(47) of \cite{BH} are not correct, due to a missing power of $y$ in the inserted $z_k^2$; in Eqs. (23)--(25) of \cite{TMB} the correct amplitude is recovered.}


\subsection{BH2}

From equations (\ref{aeffbh}) and (\ref{SigmaII}),
\be
\Sigma^2 = \frac{a_+a_-}{a^2}\,.
\ee
In the UV,
\bs 
\ba
\Sigma^2     &=& 1-\mu_*\,,\\
\sigma       &=& \frac{4\mu_*}{1-\mu_*}\,,\\
\bar{\sigma} &=& \frac{4}{1-\mu_*}\,.
\ea
\es 
For $\mu \ll 1$, $\sigma \approx 4\mu_*$ and $\bar{\sigma} \approx 4$. In the IR,
\bs 
\ba
\Sigma^2 		 &=& \frac{\delta}{\delta+1}\,,\\
\sigma 			 &=& \frac{2}{\delta+1}\,,\\
\bar{\sigma} &=& \frac{2\delta}{\delta+1}\,.
\ea
\es 
When $\delta \ll 1$, $\sigma \approx 2(1-\delta)$.


\subsection{New1}

The correction to the commutative amplitude reads
\be
\Sigma^2 = \frac{2(a_+a_-)^{3/2}}{a^2 (a_+ + a_-)}\,.
\ee
In the UV region,
\bs
\ba
\Sigma^2 		 &=& (1-\mu_*)^{3/2}\,,\\
\sigma 			 &=& \frac{6\mu_*}{1-\mu_*}\,,\\
\bar{\sigma} &=& \frac{4}{1-\mu_*}\,.
\ea
\es 
In the IR limit,
\bs 
\ba
\Sigma^2 		 &=& \left(\frac{\gamma}{1+\gamma}\right)^{3/2}\,,\\
\sigma 			 &=& \frac{6}{1+2\gamma}\,,\\
\bar{\sigma} &=& \frac{8\gamma(\gamma+1)}{(1+2\gamma)^2}\,.
\ea
\es 
In the strongly non-commutative limit ($\gamma \ll 1$), $\Sigma^2=\gamma^{3/2}$ and $\sigma = 6 +O(\delta^2)$.


\subsection{New2}

From Eqs. (\ref{aeffnew}) and (\ref{SigmaII}),
\be
\Sigma^2 = \frac{\sqrt{a_+a_-}}{a}\,.
\ee
The UV limit gives
\bs
\ba
\Sigma^2 		 &=& \sqrt{1-\mu_*}\,,\\
\sigma 			 &=& \frac{2\mu_*}{1-\mu_*}\,,\\
\bar{\sigma} &=& \frac{4}{1-\mu_*}\,.
\ea
\es
In the IR region,
\bs
\ba
\Sigma^2 		 &=& \left(\frac{\gamma}{\gamma+1}\right)^{1/2}\,,\\
\sigma 			 &=& \frac{2}{1+2\gamma}\,,\\
\bar{\sigma} &=& \frac{8\gamma(\gamma+1)}{(1+2\gamma)^2}\,.
\ea
\es
For $\gamma \ll 1$, $\sigma = 2 +O(\delta^2)$.


\section{Discussion} \label{disc}

To summarize, we can compare the considered models in the perturbative limits, that is, the UV commutative limit ($\mu \ll 1$) and the IR noncommutative limit ($\delta \approx \sqrt{\gamma} \ll 1$). Trivially, in the nonperturbative or commutative IR region ($\delta\approx\gamma \gg 1$), $a \approx a_\pm$ and one recovers the standard spectrum, $\Sigma^2=1$ and $\sigma=0$; also, by construction, the noncommutative UV region is ill-defined. 

In general, we can write the UV commutative limit of the relevant quantities as
\bs \label{deepuv}
\ba
\Sigma^2     &\approx& 1-b\mu_*\,,\\
\sigma       &\approx& 4b\mu_*\,,\label{sigapp}\\
\bar{\sigma} &\approx& 4\,,
\ea
\es
where $b$ is a constant. As anticipated, the structure of the IR amplitudes also permits a perturbative expansion around $1/\delta$; in this case, spectral amplitudes are evaluated at $k \lesssim k_0$ via the power-law solution. The IR commutative limit is then
\bs \label{uvir}
\ba
\Sigma^2     &\approx& 1-b\sqrt{\mu_*/2}\,,\\
\sigma       &\approx& 2b\sqrt{\mu_*/2}\,,\\
\bar{\sigma} &\approx& 2\,;
\ea\es 
from the previous discussions, it is natural to interpret this as an intermediate momentum region at the edge of the UV regime, around $\mu \lesssim 1$ where Eq. (\ref{aUV}) ceases to be valid, and corresponding to perturbations generated across the Hubble horizon. In fact, what one does is hit this region starting from the low-momentum IR side. The above-mentioned junction spectrum of \cite{TMB} is located somewhere closer to the infrared. 

Table \ref{table2} shows that all the models display similar asymptotic limits toward different numerical coefficients, the BH ones being larger than the New ones; the coefficient of BH1 is 4 times that of model 2 within each region (UV or IR), while this ratio is reduced to $b_1/b_2=3$ in the New model. Thus, there is less difference between model New1 and model New2 with respect to that occurring between BH1 and BH2, further confirming that the ``half-smearing'' of the new scenario somehow softens noncommutative effects.
\begin{table}[h]
\caption{\label{table2}The commutative limit, to lowest order in $\mu \ll 1$.}
\begin{ruledtabular}
\begin{tabular}{l|c}
Model  &    $b$ (commutative limit)\\\hline
BH1    &     4  \\
BH2    &     1  \\
New1   &    3/2 \\
New2   &    1/2 \\
\end{tabular}\end{ruledtabular}
\end{table}

The intermediate spectrum (\ref{uvir}) breaks down when $\Sigma^2<0$, that is when $H/M_s >$ 0.5 (BH1), 0.8 (New1), 1 (BH2) and 1.4 (New2); therefore Eq. (\ref{uvir}) well describes class 2 models at the UV boundary $\mu \lesssim 1$ while it is not particularly reliable for class 1 models.

\begin{figure}[ht]
\includegraphics[width=8.6cm]{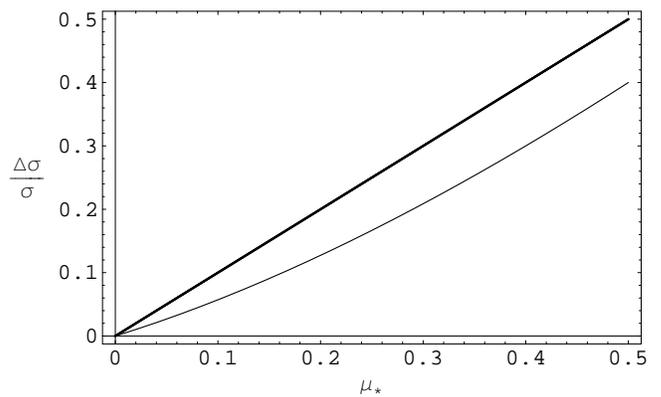}
\caption{\label{fig1} The relative approximation error $(\sigma-\sigma_\text{\tiny appr})/\sigma$ vs $\mu_*$ in the UV sector. The thin line is for BH1, the thick line is a superposition of BH2, New1 and New2.}
\end{figure}

In the deep UV or commutative limit, the linear approximation (\ref{deepuv}) properly encodes all the phenomenology of the models; however, the exact noncommutative amplitude better describes the behavior of the cosmological observables in the full span of the UV region. To see this, let us compare the function $\sigma$, governing the energy dependence of the spectral index (\ref{spindex}), with its approximated version $\sigma_\text{\tiny appr}$ given by Eq. (\ref{sigapp}); we plot the quantity $(\sigma-\sigma_\text{\tiny appr})/\sigma$ for the UV models in Fig. \ref{fig1}. The BH2, New1 and New2 models display the same linear trend in $\mu_*$, while the BH1 curve is a little below the bisector; the approximation error is up to 50\% for $\mu_* \lesssim 0.5$, correspondig to $H/M_s \lesssim 0.7$, and goes below 10\% when $H/M_s \lesssim 0.5$.
An analogous treatment of Eqs. (\ref{salph}) and (\ref{talph}) shows that the difference between the $\mu$-exact and the approximated scalar running may be even greater than the WMAP experimental error for this observable, $\alpha_s-\alpha_{s,\text{\tiny appr}} \gtrsim 10^{-2}$, for any $\theta$ and suitable values for $n_s$ and $r$ in the allowed range. Therefore, the following analysis has been conducted with the full nonlinear amplitude.

\begin{figure}[h]
\includegraphics[width=8.6cm]{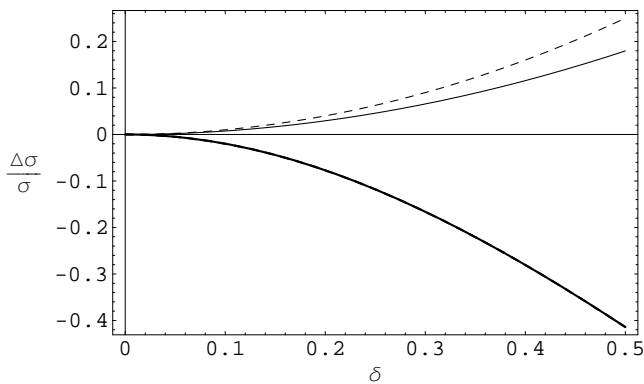}
\caption{\label{fig2} The relative approximation error $(\sigma-\sigma_\text{\tiny appr})/\sigma$ vs $\delta$ in the IR sector. The thin solid line is for BH1, the thin dashed line is for BH2 and the thick line is a superposition of New1 and New2.}
\end{figure}

Table \ref{table3} reports the noncommutative high-energy limit in the IR region. In particular, the spectral amplitude of New1 is twice the amplitude of BH1; however, within each class (1 and 2) a unique set of consistency relations is generated. In the perturbative noncommutative limit, $\delta \ll 1$, the IR version of $(\sigma-\sigma_\text{\tiny appr})/\sigma$ is shown in Fig. \ref{fig2}. The relative approximation error is up to 20\% for the BH models and $\delta \lesssim 0.5$, while it is up to 40\% for the New models. The curves of New1 and New2 models coincide.
\begin{table}[h]
\caption{\label{table3}Noncommutative zoology in the IR high-energy limit, to lowest order in $\delta \ll 1$.}
\begin{ruledtabular}
\begin{tabular}{l|cc}
Model     &\multicolumn{2}{c}{IR noncommutative limit}\\
          &   $\Sigma^2$   & $\sigma$ \\ \hline
BH1       &  $\delta^3/2$  &     6    \\
New1      &  $\delta^{3}$  &     6    \\ 
BH2       &    $\delta$    &     2    \\
New2      &    $\delta$    &     2    \\
\end{tabular}\end{ruledtabular}
\end{table}

In standard cosmology, the consistency equation relating the tensor index $n_t$ and $r$ is adopted in order to reduce the space of parameters. Until now, this has been done only for the 4D and RS cases, both displaying the same 4D degenerate version of Eq. (\ref{ntconeq}). The function $\sigma$ contains a new theoretical parameter, the string energy scale $M_s$, which enlarges the standard space of cosmological variables. In principle, this might pose some problems if one wanted a reasonably stringent constraint on the observables, facing an uncertainty similar to that one gets when keeping $n_t$ unfixed \cite{efs02}. In the UV commutative region $\sigma \ll 1$, however, one can use the known results for the 4D and RS likelihood analysis in order to compare the consistency equations in the allowed range \cite{TL}. For the Gauss-Bonnet case one should rely on the results found in \cite{DLMS,TSM}. 

The IR noncommutative limit is easier to deal with since the asymptotic form of Eq. (\ref{ntconeq}) is independent of the string scale, as it is shown in Table \ref{table4}. Some features are particularly interesting: (1) The infrared RS-2 models are the only ones with a negative tensor tilt, other noncommutative realizations giving a tilt sign opposite to that of the commutative case; (2) 4D class 2 models predict an exactly scale-invariant tensor spectrum to lowest order in SR, setting $n_t \sim O(\epsilon^2)$; (3) The highest proportionality coefficient is provided by GB class 1 models, allowing a greater tilt given the same tensor-to-scalar ratio.
\begin{table}[h]
\caption{\label{table4}The consistency equation (\ref{ntconeq}) in the commutative UV and noncommutative IR limit.}
\begin{ruledtabular}
\begin{tabular}{c|ddd}
   (Non)commutative         &\multicolumn{3}{c}{$n_t/r$}\\
      models                &       GB     &           RS    &   4D   \\ \hline
Commutative UV ($\sigma=0$) &      -1      &          -2     &   -2   \\
Class 1 IR ($\sigma=6$)     &       5      &           2     &    4   \\ 
Class 2 IR ($\sigma=2$)     &       1      &  -\frac{2}{3}   &    0   \\
\end{tabular}\end{ruledtabular}
\end{table}

Although there are $3\cdot2^4=48$ models at hand and a great amount of information to deal with, some preliminary considerations will permit us to simplify such an intricate taxonomy and draw theoretical curves in a reasonable region in the $n_s-r$ plane. Let us first compare the BH scenario with the New one and define $|\sigma|\equiv(\sigma_\text{\tiny BH}+\sigma_\text{\tiny New})/2$ and $\Delta \equiv (\sigma_\text{\tiny BH}-\sigma_\text{\tiny New})/|\sigma|$. Figure 3(a) shows that in the commutative region BH and New models are considerably different, being $\Delta_1^\text{\tiny UV}=2(\mu_*+5)/(7\mu_*+5)\sim 10/11$ when $\mu_* \rightarrow 0$ and $\Delta_2^\text{\tiny UV}= 2/3$. In the limit $\mu_* \rightarrow 1$, $\Delta_1^\text{\tiny UV} \rightarrow \Delta_2^\text{\tiny UV} $; this is a spurious effect due to the breaking of the Taylor expansion (\ref{aUV}), as one can see by considering the commutative limit of the IR spectra in Fig. 3(b). In fact, $\Delta_1^\text{\tiny IR} \neq \Delta_2^\text{\tiny IR}$ when $\delta \rightarrow \sqrt{2}$ and, as expected, $\Delta_1^\text{\tiny IR} \rightarrow 10/11$ and $\Delta_2^\text{\tiny IR} \rightarrow 2/3$ when $\delta \rightarrow \infty$. All this is in accordance with Table \ref{table2}. However, in the IR noncommutative limit there is little difference between BH and New models, being $\Delta \lesssim 10\%$. Therefore, we will only show the results of New in the infrared and skip the almost identical counterparts in BH.

A similar inspection shows that class-1 and class-2 models are quantitatively nondegenerate, getting $\sigma_1 =3\sigma_2$ for New and BH-IR, and $\sigma_1=4\sigma_2$ for BH-UV, in agreement with Tables \ref{table2} and \ref{table3}. Note that these results are independent of the bulk physics.

The versatility of the patch formalism allows coupling it to a noncommutative background in a great number of ways. For example, a realistic picture of the cosmological evolution would be to adopt one particular patch regime in a time interval when a given region of the (non)commutative spectrum is generated; one may then associate the IR region of  extra-horizon-generated perturbations with the early-Universe high-energy period, when the extra dimension opens up and the Friedmann equation suffers either GB and/or RS modifications. The consequent evolution is GB-IR $\rightarrow$ RS-IR/UV $\rightarrow$ 4D-UV. Another possibility is to consider pure energy patches and study the noncommutative spectrum in GB, RS, and 4D separately.

Let us compare the running of the scalar index of ordinary-inflaton and tachyon-inflaton fields, 
\be \label{da}
\Delta\alpha_s \equiv \alpha_s(\phi)-\alpha_s(T)\,.
\ee
Since the graphic material is very abundant, we give just a selection of it; the full set of bi- and three-dimensional figures of this and other combined analyses are available upon request to the author. In Fig. 4 the relative running $\Delta\alpha_s(n_s=1,r,\mu_*)$ is presented for 4D noncommutative models in the ultraviolet. Two-dimensional slices are then displayed in Figs. 5 and 6. Figure 5 shows that the relative running in Randall-Sundrum is rather modest; on the contrary, in GB and 4D noncommutativity may conspire to bias Eq. (\ref{da}) and, in particular, the scalar running above the current WMAP uncertainty estimates, $O(10^{-2})$. Braneworld effects, if any, should become more apparent in Planck data, for which the forecasted error is one order of magnitude smaller, $\Delta\alpha_s \sim O(10^{-3})$ \cite{BCLP}. In each 2D plot we keep the commutative model as a reference. Note that to increase either $n_s$ or $\delta$ ($\mu_*^{-1}$) pushes $\Delta\alpha_s$ toward positive values. Finally, Figs. 7 and 8 show some features of the New scenarios in the infrared region.


\section{Concluding remarks} \label{rems}

In this paper we have considered several classes of noncommutative inflationary models within an extended version of patch cosmological braneworlds, starting from a maximally invariant *-generalization of the action for scalar and tensor perturbations. Observables and consistency relations are provided via a SR approximation. The main results are:
\begin{itemize}
\item Class 1 and class 2 models are appreciably distinct from each other in the full span of the spectrum.
\item BH and New models give almost the same predictions in the IR region of the spectrum.
\item The relative running (\ref{da}) is generally more pronounced in the GB scenario than in 4D, while in RS the effect is less evident.
\item Either increasing $n_s$ or going to the commutative limit, $H/M_s\rightarrow 0$, the relative running $\Delta\alpha_s$ tends toward positive values.
\item The consistency relation $n_t \propto r$, Eq. (\ref{ntconeq}), greatly differs from one noncommutative model to another.
\end{itemize}
These models are far from being fully explored. For instance, one could impose also the extra dimension(s) to be noncommutative and extend the algebra (\ref{alg}) or other realizations to the transverse direction(s). A brane with finite thickness would emerge because of the minimum length scale $l_s$; in this case our analysis could be thought as performed on mean-valued quantities along the brane thickness. For example, $\rho \rightarrow \langle\rho\rangle \sim \int_\text{brane} \rho\, dy$, $p \rightarrow \langle p\rangle$, and so on. The subject requires further attention and a good starting point might be the cosmological thick brane setup \cite{KKOP1,KKOP2,CEHS,KKS,ML,wan02,BGS,BM,GY,KS}.

An interesting possibility is to choose another vacuum state rather than the adiabatic vacuum with which the perturbation spectrum is usually calculated. This scheme has been outlined in \cite{dan02} and developed in \cite{ACT,cai04}.

Other important aspects might be the subject of future studies. First, the use of the gravitational version of the function $z(\teta)$, Eq. (\ref{zgrav}), would permit one to compute next-to-leading-order expressions both for the tensor amplitude and the consistency equation for the tensor index \cite{cal2,cal5}. Secondly, a numerical simulation of the CMB spectrum as well as a likelihood analysis involving the consistency equation (\ref{ntconeq}), or its next-order version, in the IR limits of Table \ref{table4} are required in order to constrain the space of cosmological parameters in the low-momentum region of the perturbation spectra. Third, different analyses would point out other important aspects of the models; one may set his/her fancy free by looking at cross comparisons like in \cite{cal3} and define general relative runnings $\Delta\alpha_s \equiv \alpha_s^{(\theta)}(\psi)-\alpha_s^{(\theta')}(\psi')$. Also, stand-alone analyses with explicit inflationary models would constrain the inflaton potential according to the predictions for the cosmological observables obtained from the SR expressions for $n_t$, $n_s$, and $\alpha_s$. All these topics are currently under investigation. 


\begin{acknowledgments}
I am grateful to L. Griguolo for his invaluable suggestions and advice during the completion of this work. I also thank Y.S. Myung for useful discussions.
\end{acknowledgments}


\end{document}